\documentclass[aps,onecolumn,eqsecnum,amsmath,nofootinbib,preprintnumbers]{revtex4}%
\usepackage{float}
\usepackage{multirow}
\usepackage{mathrsfs}
\usepackage[dvips]{color,graphicx}
\usepackage{amsfonts,amssymb,theorem,mathrsfs,times}
\usepackage{amsmath}
\usepackage{amssymb}
\usepackage{color} 
\newtheorem{theorem}{Theorem}
\begin{document}

\preprint{\hfill {\small {ICTS-USTC-14-13}}}

\title{Lichnerowicz-Type Theorems for  Self-gravitating Systems with  Nonlinear Electromagnetic Fields}

\author{Li-Ming Cao$^{a,b}$\footnote{e-mail address:
caolm@ustc.edu.cn}, Yuxuan Peng$^a$\footnote{e-mail address: yxpeng@mail.ustc.edu.cn }, Jianfei Xu$^a$\footnote{e-mail address:
jfxu06@mail.ustc.edu.cn}}


\affiliation{$^a$
Interdisciplinary Center for Theoretical Study\\
University of Science and Technology of China, Hefei, Anhui 230026, China}

\affiliation{$^b$ State Key Laboratory of Theoretical Physics,
Institute of Theoretical Physics, Chinese Academy of Sciences,
P.O. Box 2735, Beijing 100190, China}

\vspace*{2.cm}
\begin{abstract}
We consider a self-gravitating system containing a globally timelike Killing vector and a nonlinear Born-Infeld electromagnetic field and  scalar fields. We prove that under certain boundary conditions (asymptotically flat/AdS) there can't be any nontrivial field configurations in the spacetime. To explore nontrivial solutions one should break any of the conditions we imposed. The case with another type of nonlinear electromagnetic field is also analyzed, and similar conclusions have been obtained under
certain conditions.
\end{abstract}
\maketitle

\section{Introduction}
AdS/CFT correspondence plays an essential role in modern
theoretical physics--- it relates a gravity theory in the bulk of an asymptotical
AdS spacetime to a  conformal
field theory without gravity on the boundary of this spacetime
\cite{Maldacena:1997re, Gubser:1998bc, Witten:1998qj}, and it develops some new methods to study  strongly coupled gauge theories
by using gravity theory which is weakly coupled. According to the dictionary of AdS/CFT correspondence, an AdS black hole
configuration corresponds to a field
theory  with a finite temperature living on the boundary of the spacetime, while a bulk asymptotically AdS gravitational system without event horizons
corresponds to a boundary field theory of zero temperature.  For instance, pure AdS spacetime is usually assumed to
correspond to the ground state of the boundary gauge theory.

However, pure (global) AdS spacetime is unstable under arbitrarily small
generic perturbations \cite{Bizon:2011gg}. This instability is
triggered by a resonant mode mixing which gives rise to the diffusion of
energy from low to high frequencies (see also \cite{Dias:2011ss}). These remarkable
results imply that pure AdS spacetime tends to form a black hole under a nonlinear perturbation.
In the framework of AdS/CFT correspondence, this also suggests that the boundary ground state always tends to thermalize. The instability of
pure AdS spacetime has inspired the study of the stability of some regular (without horizons) asymptotically AdS self-gravitating systems under nonlinear perturbations.
Some typical systems such as geons and AdS boson stars have been proved to be stable under some nonlinear perturbations~\cite{Buchel:2013uba, Maliborski:2013ula}, and, heuristically, the corresponding boundary fields have no tendency of thermalization. So it seems that the thermodynamics of  asymptotically AdS spacetimes
is more complicated than expected, and that the study of the regular self-gravitating systems is
interesting and important not only in the AdS/CFT correspondence but also in gravity side.

Recently, some numerical solutions with one Killing vector field have been found in Einstein-complex scalar
system without mass or potential \cite{Dias:2011at}. These solutions can describe some boson stars or black holes, and provide the
first example of  black hole solutions with only one Killing vector field. In the case of boson stars, the solutions are not stationary because the Killing vector field is not globally timelike and consequently some ergo surface
exists near the timelike infinity of the spacetime \cite{Dias:2011at}. A natural question we can ask is that whether there exist stationary
solutions with only one timelike Killing vector in the whole
spacetime (such a solution corresponds to a strictly stationary spacetime) with some bosonic matter.

Inspired by the solutions in \cite{Dias:2011at}, a no-go theorem or
Lichnerowicz-type theorem has been presented to answer this question
by studying some possible non-trivial self-gravitating systems
composed of p-form fields and complex scalar field without mass and
potential in strictly stationary spacetimes
\cite{Shiromizu:2012hb} (Similar discussion or generalizations can be
found in \cite{Bakon:2013mva, Shiromizu:2013tka,
Shiromizu:2013pna}). The theorem says that  the strictly stationary
spacetime that is asymptotically flat/AdS must be Minkowski
spacetime/pure AdS spacetime. Physically, the possible stationary
asymptotically AdS boson stars have been excluded by this theorem.
This no-go theorem has a close relation to the so-called
Lichnerowicz theorem (or generalized Lichnerowicz theorem). In
general relativity, the Lichnerowicz theorem tells us that the
strictly stationary vacuum spacetime must be static (see
\cite{Carter:2007} and references therein). This means that the
stationary Killing vector field is not only everywhere timelike but
also hypersurface orthogonal. One can generalize the Lichnerowicz
theorem to the case of electrovacuum, and get a generalized
Lichnerowicz theorem. Actually, this has been done in
\cite{Carter:2007}. In the system considered in~\cite{Shiromizu:2012hb}, the energy-momentum tensor of the massless
complex scalar field automatically satisfies the dominant energy
condition, so, combining the positive mass
theorem~\cite{Schoen81,Witten81,Gibbons83,Gibbons83-2}, one can show
that the spacetime must be Minkowski spacetime or pure AdS spacetime
depending on the asymptotic (boundary) conditions which have been
imposed~\cite{Shiromizu:2012hb}. Of course, Minkowski spacetime and
pure AdS spacetime are both static and in this sense the no-go
theorem proposed in~\cite{Shiromizu:2012hb} can be viewed as an
enhanced version of the Lichnerowicz theorem. However, the proof of
the Lichnerowicz theorem depends on the content of the matter fields
in the theory in general, so does the proof of the enhanced version of that. Thus it is
interesting to study theorems of this type for different matter
contents.

In this paper we present a Lichnerowicz-type
theorem  (or no-go theorem) for the non-trivial self-gravitating system composed of a
nonlinear electromagnetic field and  scalar fields all minimally
coupled to Einstein gravity. Born-Infeld theory is one, which can
eliminate the infinite classical self-energy of a charged point
particle~\cite{borinf}, of the nonlinear generalizations of Maxwell
theory. The Lagrangian density of the  Born-Infeld theory in
arbitrary dimensions is given by
\begin{equation}
\mathscr{L}_{\mathrm{BI}}=\frac{1}{b^2}(\sqrt{|\det{(g_{ab})}|}-\sqrt{|\det{(g_{ab}+b F_{ab}}}|)\,,
\end{equation}
where $F_{ab}$ is the field strength of a one-form gauge field and $b$ is a constant with the dimension of length squared. The Lagrangian density has the following
expansion in terms of the parameter $b$
\begin{equation}
\label{expansion}
\mathscr{L}_{\mathrm{BI}}=\frac{1}{b^2} \sqrt{g}\left(1-\sqrt{1+\frac{b^2}{2}
F_{ab}F^{ab}+\ldots}\right)\,,
\end{equation}
where $\sqrt{g}=\sqrt{|\det{(g_{ab})}|}$. We will focus on the
4-dimensional case and prove the Lichnerowicz-type theorem for such
a nonlinear electromagnetic field and  free scalar fields.

One can come up with another type of nonlinear electromagnetic field
with its Lagrangian in the form $-\beta(F_{ab}F^{ab})^q$, this kind of higher order terms also appears in
the low energy limit of heterotic string theory (see~\cite{Maeda09} and references therein). We also
analyze the Lichnerowicz-type theorem for this power law nonlinear
electromagnetic theory, and study the conditions under which the
theorem exists.

This paper is organized  as follows. In Sec. II, we present the
no-go theorem for Born-Infeld theory with a free real scalar field.
In Sec. III, we study the cases in which the Lagrangian for the
electromagnetic field has a power law form mentioned above. Sec. IV is devoted to
conclusions and discussion.

\section{The Lichnerowicz-type theorem for Born-Infeld/ scalar fields in 4-dimension}

\subsection{The self-gravitating system and the no-go theorem}
\label{subsec:theorem1}
We consider a spacetime $(M, g_{ab})$ with a Killing vector field $K^{a}$. Here, the manifold $M$ is assumed to be simply connected and the Killing vector field $K^{a}$ is assumed to be timelike everywhere on the manifold $M$.  In other words, the spacetime $(M, g_{ab})$ is assumed to be  strictly stationary~\cite{Carter:2007}. The action we are considering is given by
\begin{equation}
\label{Action}
S=\int d^4 x\sqrt{g}\left [\frac{1}{2}R-\Lambda+\mathcal{L}_{\mathrm{BI}}-G_{MN}(\Phi)\left(\nabla_{a}\Phi^{M}\right) \left(\nabla^{a}\Phi^{N}\right)\right]\,,
\end{equation}
where $R$ is the Ricci scalar of the spacetime, and $\Lambda$ is the cosmological constant. The Born-Infeld Lagrangian $\mathcal{L}_{\mathrm{BI}}$ has a form (We only consider the leading term of the expansion in Eq.(\ref{expansion}).)
\begin{equation}
\mathcal{L}_{\mathrm{BI}}=\frac{1}{b^2} \left(1-\sqrt{1+\frac{b^2}{2} F_{ab}F^{ab}}\right)\,,
\end{equation}
and $\Phi$ can be viewed as a harmonic mapping from the spacetime to a Riemannian manifold (target manifold) with a metric $G_{MN}$~\cite{Heusler96}.
Note that there are no potential or gauge couplings for the scalar field. Here, we have set $8\pi G=1$.

The equations of motion of the electromagnetic field has a simple form
\begin{equation}
\label{EoM}
\nabla_b \left(fF^{ab}\right)=0\, ,
\end{equation}
and the Einstein's equations can be reduced to
\begin{equation}
\label{EinsteinRicci}
R_{ab}=\frac{1}{b^2}\left(f-1\right)g_{ab}+\Lambda g_{ab}+fF_{ac}{F_{b}}^{c}+2G_{MN}\left(\nabla_{a}\Phi^{M}\right) \left(\nabla_{b}\Phi^{N}\right) \, ,
\end{equation}
where, in the above two equations, we have introduced a function $f$  which is defined by
\begin{equation}
f\equiv \Big(1+\frac{b^2}{2} F_{ab}F^{ab}\Big)^{-\frac{1}{2}}\,.
\end{equation}
The no-go theorem or Lichnerowicz-type theorem we will prove  is stated as follows:
\begin{theorem}
[Lichnerowicz-type theorem for Born-Infeld and scalar system]
If the solution to the equations of motion for the action (\ref{Action}) is strictly stationary and  asymptotically flat or asymptotically AdS, and if the dominant energy condition holds, the spacetime has to be Minkowski spacetime or pure AdS spacetime.
\end{theorem}
Actually the Born-Infeld field satisfies the dominant energy condition. Since such solutions correspond to strictly stationary spacetimes with boson field, physically the above theorem implies  these kinds of  boson stars cannot exist even theoretically.

\subsection{The proof of the theorem}
We now move on to prove the above theorem. Following the notation by Geroch in~\cite{Geroch:1970nt}, one can define the metric on the orbit space, denoted by $S$,  of the Killing field as
\begin{equation}
h_{ab}=g_{ab}-\lambda^{-1}K_{a}K_{b}\, ,
\end{equation}
where $\lambda=K_aK^a$ is always negative because the spacetime is assumed to be  strictly stationary.
The twist vector $\omega_{a}$ associated with the Killing vector field is defined as
\begin{equation}
\omega_{a}=\epsilon_{abcd}K^{b}\nabla^{c}K^{d}\, ,
\end{equation}
where $\nabla_a$ is the covariant derivative which is compatible with the metric $g_{ab}$, and $\epsilon_{abcd}$ is the natural volume element associated with $g_{ab}$ of the spacetime.  By using the properties of the Killing field, it is not hard to find~\cite{Heusler96}
\begin{equation}
\label{E.3}
\nabla_{a}\left(\lambda^{-2}\omega^{a}\right)=0\, ,
\end{equation}
\begin{equation}
\label{E.4}
\nabla_{[a}\omega_{b]}=\epsilon_{abcd}K^{d}{R^{c}}_{e}K^{e}\, ,
\end{equation}
and
\begin{equation}
\label{E.5}
\|\omega\|^2=\nabla^{a}\lambda\nabla_{a}\lambda-\lambda\nabla^{a}\nabla_{a}\lambda-2 \lambda R_{ab}K^{a}K^{b}\, .
\end{equation}
Here, $R_{ab}$ is the Ricci tensor of the spacetime. Throughout this paper we define $\|P\|^2\equiv P^{a}P_{a}$ for the orbit space vector field $P^{a}$, i.e., the vector field $P^a$ satisfies $P_aK^a=0$ and $\mathcal{L}_KP^a=0$.  By using  the timelike Killing vector field $K^a$, we can define two vectors
\begin{equation}
E_{a}=F_{ab}K^{b}\, , \qquad B_{a}=-\frac{1}{2}\epsilon_{abcd}K^{b}F^{cd}\, .
\end{equation}
Obviously, $E_{a}K^{a}=B_{a}K^{a}=0$. We also require that all possible fields on the spacetime are invariant under the transformation generated by $K^{a}$, and this suggests $\mathcal{L}_{K}F^{ab}=0$, $\mathcal{L}_{K}E_{a}= \mathcal{L}_{K}B_{a}=0$, and $\mathcal{L}_{K}\Phi^{M}=0$.  All of these indicate that $\Phi^M$, $E_{a}$, and $B_{a}$ are scalar and vectors on the orbit space $S$. Furthermore, the field $F_{ab}$ can be written as
\begin{equation}
F_{ab}=-\lambda^{-1}\left(2K_{[a}E_{b]}+\epsilon_{abcd}K^{c}B^{d}\right)\, .
\end{equation}

With the definitions above, by projecting Eq.(\ref{EoM}) along $K^{a}$, we have
\begin{equation}
\label{E.8}
\lambda\nabla_{a}\left(\lambda^{-1}fE^{a}\right)-\lambda^{-1}f\omega^{a}B_{a}=0\, ,
\end{equation}
while the projection of Eq.(\ref{EoM}) along $h_{ab}$ gives
\begin{equation}
\label{E.9}
2\nabla_{[a}\left(fB_{b]}\right)=0\, .
\end{equation}
While deriving Eq.(\ref{E.9}), we consider the equation $h_{ab}\nabla_{c}\left(fF^{ac}\right)=0.$
The left hand side is
\begin{equation}
\begin{split}
h_{ab}\nabla_{c}\left(fF^{ac}\right)
&=h_{ab}\nabla_{c}fF^{ac}+fh_{ab}\nabla_{c}F^{ac}\, ,
\end{split}
\end{equation}
in which the first term is
\begin{equation}
\begin{split}
-\lambda^{-1}\nabla_{c}f h_{ab}\left(2K^{[a}E^{c]}+\epsilon^{acde} K_{d}B_{e}\right)
&=-\lambda^{-1}\nabla_{c}f \left(-K^{c}E_{b}+g_{ab}\epsilon^{acde} K_{d}B_{e}\right)\\
&=-\lambda^{-1}\nabla_{c}f g_{ab}\epsilon^{acde} K_{d}B_{e}\, ,
\end{split}
\end{equation}
where the stationarity condition of $f$ is used. Contracting this object with $\epsilon^{aghb}K_{h}$ gives
\begin{equation}
\begin{split}
-\lambda^{-1}\epsilon^{aghb}\epsilon_{bcde}\nabla^{c}f K^{d}B^{e}
&=-3!\lambda^{-1}\nabla^{[a}fK^{g}B^{h]}K_{h}\\
&=2\nabla^{[a}fB^{g]}\, .
\end{split}
\end{equation}
Then consider the second term. Using the stationarity conditions of the fields $E_a$ and $K_a$ and the fact that $K^{a}$ is a Killing vector, one finds the second term gives $-fh_{fb}\epsilon^{fcde}\nabla_c\left(\lambda^{-1}K_d B_e\right).$ Once more, contracting this object with $\epsilon^{aghb}K_{h}$ gives
\begin{equation}
\begin{split}
&\quad-f\epsilon^{aghb}K_h{h^{f}}_{b}\epsilon_{fcde}\nabla^c \left(\lambda^{-1}K^d B^e\right)\\
&=-3!fK_h\nabla^{[a}\left(\lambda^{-1}K^g B^{h]}\right)\\
&=2f\nabla^{[a}B^{g]}\, .
\end{split}
\end{equation}
So far it is easy to see that combining the two terms give
\begin{equation}
2\nabla_{[a}\left(fB_{b]}\right)=0\, .
\end{equation}
We get, from the Bianchi identity, the equations below:
\begin{equation}
\label{E.10}
\lambda\nabla_{a}\left(\lambda^{-1}B^{a}\right)+\lambda^{-1}\omega^{a}E_{a}=0\, ,
\end{equation}
and
\begin{equation}
\label{E.11}
2\nabla_{[a}E_{b]}=0\, .
\end{equation}
As to Eq.(\ref{E.11}), it is obtained by projecting the Bianchi identity $\nabla_{[a}F_{bc]}=0$ to the $K^a$ direction:
\begin{equation}
\begin{split}
3K^a\nabla_{[a}F_{bc]}&=-\lambda^{-1}\epsilon_{bcde}K^a \nabla_a K^d B^e-\lambda^{-1}\epsilon_{bcde}K^a \nabla_a B^e K^d+\nabla_b E_c\\
&\quad-\lambda^{-1}\epsilon_{cade}K^a \nabla_b K^d B^e-\nabla_c E_b-\lambda^{-1}\epsilon_{abde}K^a \nabla_c K^d B^e\\
&=2\nabla_{[b} E_{c]}\\
&=0\, .
\end{split}
\end{equation}
The fact that the $\epsilon$ terms cancel each other can be seen by taking a Hodge dual.

On the other hand, from the Einstein's equations (\ref{EinsteinRicci}) one can get
\begin{equation}
\label{E.12}
R_{ab}K^{a}K^{b}=\frac{\lambda}{b^2}\left(f-1\right)+\lambda~\Lambda+f\|E\|^2\, ,
\end{equation}
and
\begin{equation}
\label{E.13}
R_{ab}K^{a}{h_{c}}^{b}=-\lambda^{-1} f\epsilon_{cabd}E^{a}B^{b}K^{d}\, .
\end{equation}
In the above calculation, we have used the  stationary condition of the scalar field $\Phi^M$. Considering Eq.(\ref{E.5}) and Eq.(\ref{E.12}), we get
\begin{equation}
\label{E.20}
\lambda^{-2}\|\omega\|^2=-\nabla^{a}\left(\lambda^{-1}\nabla_{a}\lambda\right)-2\left[\frac{1}{b^2}\left(f-1\right)+\Lambda+\lambda^{-1}f\|E\|^2\right]\, ,
\end{equation}
while Eq. (\ref{E.4}) and Eq.(\ref{E.13}) together give a relation
\begin{equation}
\label{E.21}
\nabla_{[a}\omega_{b]}=2fE_{[a}B_{b]}\, .
\end{equation}

Because of Eq. (\ref{E.11}) we can introduce a function $E$ so that $E_{a}=-\nabla_{a}E$ at least locally. Furthermore, the definition of $E$ can be extended on the whole spacetime because the manifold $M$ has been assumed to be simply connected. Substituting the $E_{a}$ into Eq. (\ref{E.21}) and using the result (\ref{E.9}), we get
\begin{equation}
\nabla_{[a}\left(\omega_{b]}+2f E B_{b]}\right)=0\, .
\end{equation}
Then we can introduce a function $\psi$  so that
\begin{equation}
\omega_{a}+2f E B_{a}=2\nabla_{a}\psi\, .
\end{equation}
By contracting both sides in the above equation with $\lambda^{-2}\omega^{a}$ and using Eq. (\ref{E.3}), we find
\begin{equation}
\label{psi}
\nabla_{a}\left(2\lambda^{-2}\omega^{a}\psi\right)=\lambda^{-2}\|\omega\|^2+2\lambda^{-2}f E B^{a}\omega_{a}\, .
\end{equation}
At the same time, Eq.(\ref{E.8}) implies that
\begin{equation}
\label{E}
\nabla_{a}(\lambda^{-1} fEE^{a})=-\lambda^{-1}f\|E\|^2+\lambda^{-2}fE B^{a}\omega_{a}\, .
\end{equation}
The above two equations give
\begin{equation}
\label{psi&E}
\nabla_{a}\left(2\lambda^{-2}\omega^{a}\psi-2\lambda^{-1} f E E^{a}\right)=\lambda^{-2}\|\omega\|^2+2\lambda^{-1}f\|E\|^2\, .
\end{equation}
Using one more time the above method --- introducing the function $B$ so that $fB_{a}=\nabla_{a}B$ and substituting the $B_{a}$ into Eq. (\ref{E.21}) --- we get for some function $\phi$
\begin{equation}
\omega_{a}+2B E_{a}=2\nabla_{a}\phi\, ,
\end{equation}
and
\begin{equation}
\label{phi}
\nabla_{a}\left(2\lambda^{-2}\omega^{a}\phi\right)=\lambda^{-2}\|\omega\|^2+2\lambda^{-2}B E^{a}\omega_{a}\, .
\end{equation}
Meanwhile, Eq. (\ref{E.10}) indicates that
\begin{equation}
\label{B}
\nabla_{a}(\lambda^{-1}B B^{a})=\lambda^{-1}f\|B\|^2-\lambda^{-2}B E^{a}\omega_{a}\, .
\end{equation}
The above two equations give a relation
\begin{equation}
\label{phi&B}
\nabla_{a}\left(2\lambda^{-2}\omega^{a}\phi+2\lambda^{-1}B B^{a}\right)=\lambda^{-2}\|\omega\|^2+2\lambda^{-1}f\|B\|^2\, .
\end{equation}
Combining Eqs. (\ref{psi&E}) and (\ref{phi&B}) and using Eq. (\ref{E.20}), we get
\begin{eqnarray}
\label{ADM}
&&\nabla_{a}\left[\lambda^{-1}\nabla^{a}\lambda+\lambda^{-2}\omega^{a}\left(\psi+\phi\right)+ \lambda^{-1}\left(B B^{a}-f E E^{a}\right)\right]+2\Lambda\nonumber\\
&&=-\frac{2}{b^2}\left(f-1\right)-\lambda^{-1}f\|E\|^2+\lambda^{-1}f\|B\|^2\nonumber\\
&&=-\frac{f}{b^2}\left(f^{-1}-1\right)^2.
\end{eqnarray}

This equation is the key to our proof.

Firstly, let us consider the case with $\Lambda=0$.  When the spacetime is asymptotically flat, following the method by~\cite{Shiromizu:2012hb}, the space volume integral of the left hand side of the above equation  can be turned into a 2-surface integral at infinity. The contribution from the first term $\nabla_{a}\left(\lambda^{-1}\nabla^{a}\lambda\right)$ is proportional to the total mass of the whole spacetime, and the other two terms give no contribution since they diminish as or faster than inverse of the square of the radial coordinate at infinity~\cite{Shiromizu:2012hb}. But apparently the right hand side will never be positive. So according to the positive mass theorem, as long as the dominant energy condition holds, the total mass must be zero. (Actually the Born-Infeld field satisfies the dominant energy condition, which can be seen by contracting the energy-momentum tensor with any timelike vector field.) Then the spacetime must be Minkowski spacetime~\cite{{Schoen81},{Witten81}}. 

Secondly, we consider  the case with $\Lambda<0$. When the spacetime is asymptotically AdS, again, following the method by~\cite{Shiromizu:2012hb}, we can introduce the vector field $r^{a}$ which satisfies
\begin{equation}
\nabla_{a}r^{a}=-2\Lambda\, ,
\end{equation}
then Eq.(\ref{ADM}) becomes
\begin{equation}
\nabla_{a}\left[\lambda^{-1}\nabla^{a}\lambda+\lambda^{-2}\omega^{a}\left(\psi+\phi\right)+ \lambda^{-1}\left(B B^{a}- fEE^{a}\right)-r^{a}\right]
=-\frac{f}{b^2}\left(f^{-1}-1\right)^2\, .
\end{equation}
The left hand side, after integration, is again a surface integral at infinity and proportional to the total mass of the spacetime~\cite{Shiromizu:2012hb}. We use the positive mass theorem for asymptotically AdS spacetimes and conclude that the spacetime must be pure AdS spacetime~\cite{{Gibbons83},{Gibbons83-2}}.

The theorem proposed in
sec.(\ref{subsec:theorem1}) has been proven in the above discussion.

\section{Another type of nonlinear gauge field}
So far it has been proven that in the above cases there can't be any nontrivial field configurations. This is the main conclusion of this paper. But one can't help ask if this conclusion still holds when other types of nonlinear electrodynamics are considered, and furthermore how universal this phenomenon (the no-go theorem) is. An example of other kinds of electrodynamics is given below:
\begin{equation}
\mathcal{L}_{\mathrm{EM}}=-\beta\left(F_{ab}F^{ab}\right)^q\, ,
\end{equation}
where $\beta$ is a coupling constant and $q$ is a rational number. (This kind of nonlinear term of gauge field can be found in
the low energy limit of heterotic string theory, see reference~\cite{Maeda09} in which  the black hole solutions for this model in Lovelock gravity theory have been studied.)
Replacing the electromagnetic field in Eq.(\ref{Action}) with the above one while keeping other parts of the system unchanged we obtain the equations of motion below:
\begin{equation}
R_{ab}=\tilde{f}F_{ac}{F_{b}}^{c}+\beta\left(F^2\right)^q\left(1-2q\right) g_{ab}+\Lambda g_{ab}+2G_{MN}\left(\nabla_{a}\Phi^{M}\right) \left(\nabla_{b}\Phi^{N}\right)\, ,
\end{equation}
and
\begin{equation}
\nabla_{b}\left(\tilde{f} F^{ab}\right)=0\, ,
\end{equation}
where
\begin{equation}
F^2\equiv F_{ab}F^{ab}\, ,
\end{equation}
and
\begin{equation}
\tilde{f}\equiv 4\beta q\left(F^2\right)^{q-1}\, .
\end{equation}
For such a system we can apply the same method in the last section and the only differences from the case of the Born-Infeld field are in the Einstein's equations as long as we keep the notation $\tilde{f}$ during the computation. Finally we get the following equation as the counterpart of Eq.(\ref{ADM})
\begin{equation}
\label{Power law ADM}
\nabla_{a}\left[\lambda^{-1}\nabla^{a}\lambda+\lambda^{-2}\omega^{a}\left(\psi+\phi\right)+ \lambda^{-1}\left(B B^{a}- \tilde{f}EE^{a}\right)\right]+2\Lambda
=-2\beta\left(F^2\right)^q\left(1-q\right)\, .
\end{equation}
If the right hand side is again zero or negative, we would have shown that for such a system there can't be nontrivial field configurations. But obviously there are chances for it to be positive.

However, there are other constraints which can be derived from the dominant energy condition, which is a prerequisite while using the positive mass theorem to prove the no-go theorem. We will find out these constraints soon in certain cases, and see if the right hand side of Eq.(\ref{Power law ADM}) will be restricted to a non-positive quantity, thus leading to a no-go theorem.

\subsection{The electric field case}
First consider the case in which there are no fields other than the electric field and the scalar field in the point of view of the some observers. This means that if we carefully choose some  orthogonal basis $\left\{(e_0)^{a},(e_1)^a,(e_2)^a,(e_3)^a\right\}$  only the entries $F_{0i}$ ($i=1,2,3$) of the component matrix $F_{\mu\nu}$ are nonzero. Furthermore, $F^2$ is negative. Without loss of generality, we can choose the vectors $(e_1)^a,(e_2)^a,(e_3)^a$ in order that only $F_{01}=-F_{10}\neq 0$.
Under this basis and the corresponding dual basis, the energy momentum tensor of the electromagnetic field has its component matrix diagonal:
\begin{equation}
  \left\{
   \begin{aligned}
   {T^{i}}_{0}&=0,  \\
   {T^{i}}_{j}&=0, \qquad (i,j=1,2,3; \quad i\neq j) \\
   \end{aligned}
   \right.
  \end{equation}
since
\begin{equation}
{T^{\mu}}_{\nu}=4\beta q (F^2)^{q-1}F^{\mu\rho}F_{\nu\rho}-\beta{\delta^{\mu}}_{\nu}(F^2)^{q}\, ,
\end{equation}
and the dominant energy condition can be expressed as
\begin{equation}
-{T^{0}}_{0}\geqslant \left|{T^{i}}_{i}\right| \qquad \text{(no sum)}.
\end{equation}
The above inequations give that
\begin{equation}
  \left\{
   \begin{aligned}
   &\beta\left(1-2q\right)(F^2)^{q}\geqslant 0\, ,  \\
   &-\beta\left(1-2q\right)(F^2)^{q}\leqslant -\beta(F^2)^{q}\leqslant \beta\left(1-2q\right)(F^2)^{q}\, . \\
   \end{aligned}
   \right.
  \end{equation}
These conditions will have restrictions on the parameters $\beta$ and $q$. We find that when $\beta>0$, it is required that if $q\geqslant 1$, $q$ is an odd integer or a rational number with both the denominator and the numerator odd, and if $q< 0$, $q$ is an even integer or a rational number with the denominator odd and the numerator even; when $\beta<0$ it is required that if $q\geqslant 1$, $q$ is an even integer or a rational number with the denominator odd and the numerator even, and if $q< 0$, $q$ is an odd integer or a rational number with both the denominator and the numerator odd. The constraints are shown in the table below:
\begin{table}[H]
\centering

\begin{tabular}{|c|c|}
\hline
\multirow{2}{*}{$\beta>0$} & $q\geqslant 1$, and $q$ is an odd integer or a rational number with both the denominator and the numerator odd \\
\cline{2-2}
& $q< 0$, and $q$ is an even integer or a rational number with the denominator odd and the numerator even \\
\hline
\multirow{2}{*}{$\beta<0$} & $q\geqslant 1$, and $q$ is an even integer or a rational number with the denominator odd and the numerator even\ \\
\cline{2-2}
& $q< 0$, and $q$ is an odd integer or a rational number with both the denominator and the numerator odd \\
\hline
\end{tabular}
\caption{\label{electric case} The dominant energy condition in the electric field case}
\end{table}

Under these conditions, it is easy to find that the right hand side of Eq.(\ref{Power law ADM}), $-2\beta\left(F^2\right)^q\left(1-q\right)$, is always non-positive. We obtain the same conclusion as in the Born-Infeld case: there can't be any nontrivial field configurations.

\subsection{The magnetic field case}
In this case, there are no fields other than the magnetic field and the scalar field in the point of view of some observers, and $F^2$ is positive. Using a similar tetrad as above, diagonalizing ${T^{\mu}}_{\nu}$ and we find that the dominant energy condition requires that
\begin{equation}
\left\{
   \begin{aligned}
   &\beta>0,  \\
   &0<q\leqslant 1\, . \\
   \end{aligned}
   \right.
\end{equation}
The right hand side of Eq.(\ref{Power law ADM}) is again non-positive under this condition. So in the case of magnetic field there can't be nontrivial field configurations either.

In a word, in the above two cases (the electric field case and the magnetic field case), as long as the dominant energy condition holds, there can't be any nontrivial field configurations.

\section{conclusions and discussion}
In this paper we have proved that for some nonlinear electrodynamics the strictly stationary spacetime must be trivial, as long as the boundary conditions and the form of the couplings are fixed, just like in the case of Maxwell field~\cite{Shiromizu:2012hb}. If one wishes to explore a new solution, for example a bonson star, except for breaking any of the conditions imposed, one has to consider different models with different fields or couplings.

Actually if we introduce the gauge coupling between the electromagnetic field and the scalar fields or some potential for the scalar fields, we can't get a conclusion that the total mass is vanishing, at least by using the method in this paper. So if there are some sort of interactions (not restricted to gauge couplings) nontrivial solutions might exist. However, one should be careful while choosing the form of the interaction. For example, it's easy to find that the Lagrangian containing a dilaton-like coupling (see~\cite{Allaverdizadeh:2014xaa} for a five-dimensional case)
\begin{equation}
\mathcal{L}=\frac{1}{2}R-\Lambda+\frac{A(\Phi)}{b^2} \left(1-\sqrt{1+\frac{b^2}{2} B(\Phi) F_{ab}F^{ab}}\right)-\frac{1}{2}\nabla_{a}\Phi\nabla^{a}\Phi
\end{equation}
with $A(\Phi)$ and $B(\Phi)$ arbitrary functions of the scalar field and $A(\Phi)$ positive will give the same result as before.

It is also viable to consider modified gravity theories other than the general relativity. Recently a similar no-go theorem in Einstein-Gauss-Bonnet gravity in strictly static vacuum has been discussed in \cite{Shiromizu:2013tka}, the method in which is slightly  different from that in this paper. One might be curious if the theorem can be generalized to the cases with matter or strictly stationary cases in this modified gravity.
Changing the scalar field or introducing non-minimal couplings between matter and gravity may also be ways to find solutions, while it's possible that similar no-go theorems can be proved for these cases and can rule out some solutions. This is left for future work.

\section{Acknowledgments}

This work was supported in part by the National Natural Science Foundation of China with grants  No.11205148 and No.11235010.

\end{document}